# Limit of *zT* enhancement in rock-salt structured chalcogenides by band convergence?

Min Hong,[1] Zhi-Gang Chen,[1,]* Yanzhong Pei,[2] Lei Yang,[1] and Jin Zou[1,3,]*

[1]*Materials Engineering, University of Queensland, Brisbane, Queensland 4072, Australia*

[2]*Materials Science and Engineering, Tongji University, Shanghai 201804, China*

[3]*Centre for Microscopy and Microanalysis, University of Queensland, Brisbane, Queensland 4072, Australia*

*Corresponding author: j.zou@uq.edu.au, and z.chen1@uq.edu.au

Rock-salt structured chalcogenides, such as PbTe, PbSe, and SnTe, are the top candidates for mid-temperature thermoelectric applications, and their p-type thermoelectric efficiencies can be enhanced via aligning the valence bands. Here, we provided comprehensive numerical investigations on the effects of band convergence on electronic properties. We found that the extra valance band can indeed significantly enhance the power factor. Nevertheless, the extra valance band can also increase the electronic thermal conductivity, which partially offsets the enhanced power factor for the overall figure-of-merit. Finally, we predicted that the maximum figure-of-merit for PbTe, PbSe, and SnTe can reach to 2.2, 1.8, and 1.6, respectively, without relying on the reduction in lattice thermal conductivity.

Thermoelectricity enables the direct conversion between heat and electricity, offering a sustainable green energy technique for power generation or refrigeration [1,2]. To realize the wide applications, extensive strategies have been dedicated to enhancing the conversion efficiency, gauged by the figure-of-merit ($zT$), which can be expressed as $zT = S^2\sigma T/\kappa$, where $S$, $\sigma$, $\kappa$, and $T$ are the Seebeck coefficient, electrical conductivity, thermal conductivity (including electronic $\kappa_e$, lattice $\kappa_l$, and bipolar $\kappa_{bi}$ components), and the working temperature, respectively [3]. Among them, band engineering is widely used to tune the electronic band structures to pursue high power factor ($S^2\sigma$) [4-6], and other is to enhance phonon scatterings to reduce $\kappa$ by involving different phonon scatterings [7,8].

As dominating candidates working at mid-temperature range, rock-salt structured chalcogenides, such as PbTe, PbSe, and SnTe, have been paid extensive attentions [9-13]. They share the similar band structures, in which two extrema at the L ($E_{VL}$) and Σ ($E_{VΣ}$) points of the Brillouin zone are separated by an energy bias ($\Delta E = E_{VL} - E_{VΣ}$), which is comparable to the band gap ($E_g = E_C - E_{VL}$, with $E_C$ denoting the extreme of conduction band) [14]. Since the Σ valance band (VB$_Σ$) locates further away from the Fermi level ($E_f$) compared with the L valance band (VB$_L$), $S$ tensor of VB$_Σ$ is larger than that of VB$_L$ [15-17]. Besides, the VB$_Σ$ band degeneracy ($N_Σ$) of rock-salt structured chalcogenides is 12, much larger than the VB$_L$ band degeneracy ($N_L = 4$) [18]. In this regard, producing the convergence of VB$_L$ and VB$_Σ$ (*i.e.* reducing $\Delta E$) may greatly enhance the thermoelectric performance, when the doping is properly tuned. Experimentally, forming PbTe$_{1-x}$Se$_x$ alloys can align VB$_L$ and VB$_Σ$, which leads to $zT$ up to 1.8 [19,20]. Sr doping was employed to reduce the $\Delta E$ for PbSe [21]. On the other hand, Mn [22,23], Cd [24,25], and Hg [12] were successfully used to reduce the $\Delta E$ for SnTe. In both PbSe and SnTe with reduced $\Delta E$, $zT$ values were significantly enhanced.

Despite these great achievements, there still exist several theoretical issues that need to be fully examined. First, the band convergence temperature ($T_{cvg}$) for achieving the maximum $zT$ has not been fully clarified. For example, PbTe with $T_{cvg}$ of ~450 K [21], alloying with Se to increase $T_{cvg}$ [19,20] and doping with Mn to decrease $T_{cvg}$ [26,27] can all increase $S^2\sigma$. Second, increasing the contribution from VB$_Σ$ can increase $\kappa$. On one hand, increasing the contribution from VB$_Σ$ leads to large $\sigma$, and therefore inevitably increase $\kappa_e$. On the other hand, additional heat flow is always generated during the electron transition between VB$_L$ and VB$_Σ$ [28]. Last but not least, the determination of optimal Hall carrier concentration ($n_H^{opt}$) for maximizing $S^2\sigma$ and $zT$ is in great demand for achieving the maximum thermoelectric efficiency in multi-band situations [4,21].

In this study, we used a three-band (CB, VB$_L$, and VB$_Σ$) model to numerically investigate the impact of band convergence on tailoring thermoelectric performance in rock-salt structured chalcogenides. Detailed equations for calculating thermoelectric properties are presented in Supplemental Material, including the tensors of CB (with subscript of C), VB$_L$ (with subscript of L), and VB$_Σ$ (with subscript of Σ), and parameters used in our calculations are listed in TABLE SI. On this basis, we simulated the variations of thermoelectric properties with reduced Fermi level ($\eta$) for SnTe, as an example, over a wide temperature rang. We found that for maximizing $S^2\sigma$ and therefore $zT$, $T_{cvg}$ should equal to the highest working temperature, which is about 900 K for SnTe, PbTe, and PbSe. Through producing band convergence to enhance $S^2\sigma$, $\kappa$ is also increased, which partially offsets the enhancement in $S^2\sigma$ for the overall $zT$. In addition, we investigated $n_H$ dependent thermoelectric properties and determined the temperature dependent $n_H^{opt}$ for $S^2\sigma$ and $zT$ in PbTe, PbSe, and SnTe, respectively. Using the reported $\kappa_l$ values for these rock-salt structured chalcogenides from literatures [10,14,29], we predicted the maximum $zT$ to be 2.2, 1.8, and 1.6 for PbTe, PbSe, and SnTe, respectively. If the $\kappa_l$ reaches to the amorphous limit, $zT$ could be further enhanced to 3.1, 2.4, and 2.2 for PbTe, PbSe, and SnTe, respectively. This study suggests that there are still rooms for the $zT$ enhancement in these rock-salt structured chalcogenides by producing band convergence at 900 K and appropriately tuning $n_H$.

To quantitatively understand the contribution of $VB_\Sigma$ on thermoelectric properties, we used SnTe as an example. According to Eqs. (S1) - (S8), for a given material at a certain temperature, its thermoelectric properties vary with $\eta$ [9,30,31]. Thus, we calculated the thermoelectric properties as a function of $\eta$ at a temperature range of 300 – 900 K. The calculated results are shown in the videos of Supplemental Material, among which we highlighted $S^2\sigma$. Noteworthy, the modeling here does not taking into account the possible additional carrier scatterings by various defects simultaneously introduced during either engineering the band, optimizing the carrier concentration or lowing the $\kappa_l$. If these do exist, the model might overestimate $S^2\sigma$ and $zT$.

Fig. 1(a) shows the temperature-dependent $S^2\sigma$ as a function of $\eta$, in which two peaks can be found at low temperature. With increasing $T$, they converge into one peak (*e.g.* > 700 K) to achieve a higher $S^2\sigma$ of ~2.1×10$^{-3}$ Wm$^{-1}$K$^{-2}$ at 900 K. In the calculations, we used $\Delta E = 0.45-2.5\times10^{-4}T$ [32], which means $\Delta E = 0.22$ eV at 900 K. Despite of such inherent temperature-dependent $\Delta E$, we calculated $S^2\sigma$ for $\Delta E = 0$ eV at 900 K and for $VB_\Sigma$ overtaking $VB_L$ to be the primary valance band (*i.e.* $\Delta E < 0$ eV). As can be seen, the peak $S^2\sigma$ increases to ~3×10$^{-3}$ Wm$^{-1}$K$^{-2}$ for $\Delta E = 0$ eV at 900 K, while decreases with further reducing $\Delta E$, for instance, to be -0.1 eV. Therefore, to maximize the peak $S^2\sigma$ in the situation of multi bands, the band convergence should occur at highest working temperature.

As a comparison, we calculated $S^2_{CL}\sigma_{CL}$ by only considering $VB_L$ and CB. Fig. 1(b) shows the temperature-dependent $S^2_{CL}\sigma_{CL}$ as a function of $\eta$, in which the $S^2_{CL}\sigma_{CL}$ peaks stabilize at $\eta \approx 0.3$, agreed with the previous study [33]. Moreover, the magnitude of the $S^2_{CL}\sigma_{CL}$ peak decreases with increasing $T$, due to (1) the stronger bipolar effect at high temperature (in turn reducing the thermoelectric performance); and (2) the increased effective mass for SnTe at high temperature [14], because large effective mass reduces $S^2_{CL}\sigma_{CL}$ [9].

Through subtracting $S^2_{CL}\sigma_{CL}$ from $S^2\sigma$, we can evaluate the contribution from $VB_\Sigma$. Fig. 1(c) exhibits the calculated $S^2\sigma - S^2_{CL}\sigma_{CL}$ as a function of $\eta$. With increasing $T$, the peak of $S^2\sigma - S^2_{CL}\sigma_{CL}$ shifts to low $\eta$, and the magnitude of the peak increases. To understand this, Fig. 1(d) plots the temperature-dependent weighted mobility ratio between $VB_L$ and $VB_\Sigma$ ($B_{L/\Sigma}$, refer to Eq. (S24) for its calculation) and $\Delta E$, in which both $B_{L/\Sigma}$ and $\Delta E$ decrease with increasing $T$. According to the discussion in Section 3 of Supplemental Material, decreasing $B_{L/\Sigma}$ and $\Delta E$ lead to an enhanced contribution from $VB_\Sigma$ in $S^2\sigma$ (corresponding to the increased peak magnitude of $S^2\sigma - S^2_{CL}\sigma_{CL}$ at high temperature). The shift of $S^2\sigma - S^2_{CL}\sigma_{CL}$ peak to low $\eta$ at high temperature is caused by the reduced $\Delta E$ (*i.e.* $VB_\Sigma$ moves close to $VB_L$).

In addition, the larger difference of peak positions between $S^2_{CL}\sigma_{CL}$ and $S^2\sigma - S^2_{CL}\sigma_{CL}$ at low temperature suggests that, at low temperature, $S^2\sigma$ is mainly contributed by CB+$VB_L$ (*i.e.* $S^2_{CL}\sigma_{CL}$) at low $\eta$, but is mainly contributed by $VB_\Sigma$ (*i.e.* $S^2\sigma - S^2_{CL}\sigma_{CL}$) at high $\eta$. From Fig. 1(a), for $T \leq 700$ K, $S^2\sigma$ has two peaks and the maximum value of $S^2\sigma$ corresponds to the peak at low $\eta$. Since the $S^2_{CL}\sigma_{CL}$ peak decreases with increasing $T$ (refer to FIG 1b), the maximum value of $S^2\sigma$ at low $\eta$ decreases accordingly. With increasing $T$, the difference of peak positions between $S^2_{CL}\sigma_{CL}$ (refer to Fig. 1(b)) and $S^2\sigma - S^2_{CL}\sigma_{CL}$ (refer to Fig. 1(c)) becomes smaller, leading to the overlap of the two $S^2\sigma$ peaks at high temperature (refer to Fig. 1(a)). Furthermore, with increasing $T$, $S^2_{CL}\sigma_{CL}$ peak reduces while $S^2\sigma - S^2_{CL}\sigma_{CL}$ peak increases, so that the contribution of $VB_\Sigma$ on $S^2\sigma$ becomes significant at high temperature.

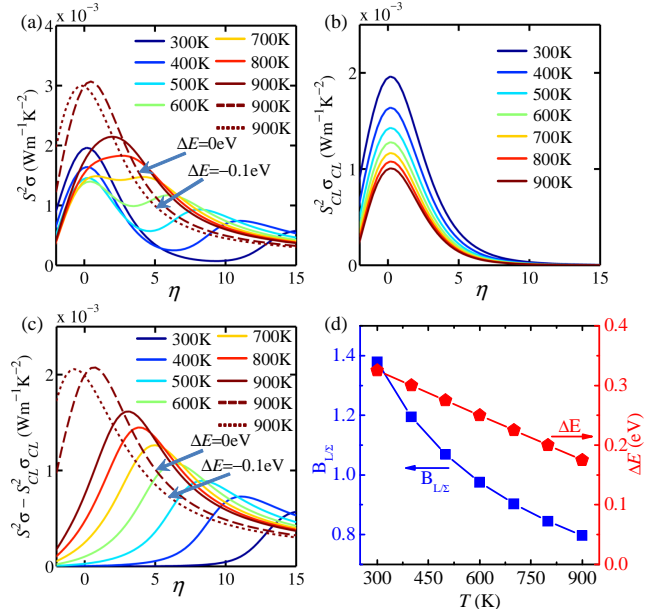

FIG. 1 Calculated (a) $S^2\sigma$, (b) $S^2_{CL}\sigma_{CL}$, and (c) $S^2\sigma - S^2_{CL}\sigma_{CL}$, as a function of $\eta$ for SnTe at temperature ranging from 300 – 900 K, respectively. (d) Temperature dependent $B_{L/\Sigma}$ and $\Delta E$ for SnTe.

In order to calculate $zT$, we should determine $\kappa$ first, including the components of $\kappa_l$, $\kappa_e$, and $\kappa_{bi}$ [34]. Herein, we calculated $\kappa_e$ and $\kappa_{bi}$ based on Eqs. (S7) and (S15), respectively, while $\kappa_l$ for SnTe was obtained from Ref. [14]. Fig. 2(a) shows the calculated $\kappa_e$ as a function of $\eta$, in which $\kappa_e$ increases with increasing $T$ and $\eta$. This is because $\kappa_e$ is the thermal energy transported by free charger carriers [35], high $T$ causes higher average thermal energy transported by individual free charger carriers, resulting in a high $\kappa_e$. In addition, free charger carrier concentration increases with increasing $\eta$ [36], large $\eta$ can lead to high $\kappa_e$. Moreover, we also calculated $\kappa_e$ for $\Delta E = 0$ and -0.1 eV, as an example, at 900 K, respectively. As can be seen, with decreasing $\Delta E$, $\kappa_e$ increases, which could partially offset the enhancement in $S^2\sigma$ caused by the decreased $\Delta E$.

Fig. 2(b) shows the calculated temperature-dependent $\kappa_{bi}$ as a function of $\eta$. Since $S^2\sigma$ reaches the peak value at 900 K and $\eta \approx 1$ (refer to Fig. 1(a)), we can examine the corre-sponding $\kappa_e$ and $\kappa_{bi}$ in this case. From Figs. 2(a) and 2(b), we found $\kappa_e \approx 0.5$ Wm$^{-1}$K$^{-1}$ and $\kappa_{bi} \approx 0.2$ Wm$^{-1}$K$^{-1}$ at 900 K and $\eta \approx 1$. Interestingly, in this case, $\kappa_{bi}$ is roughly 25% of $\kappa_e$, suggesting that $\kappa_{bi}$ plays an important role for the overall $zT$. Moreover, at 900 K, with reducing $\Delta E$ to 0 eV and even to -0.1 eV, $\kappa_{bi}$ was favorably decreased significantly; suggesting band convergence can also suppress bipolar conduction.

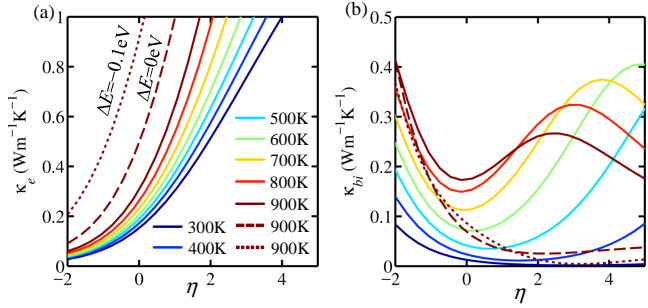

FIG. 2 Calculated (a) $\kappa_e$ and (b) $\kappa_{bi}$ as a function of $\eta$ for SnTe at temperature ranging from 300 – 900 K, respectively.

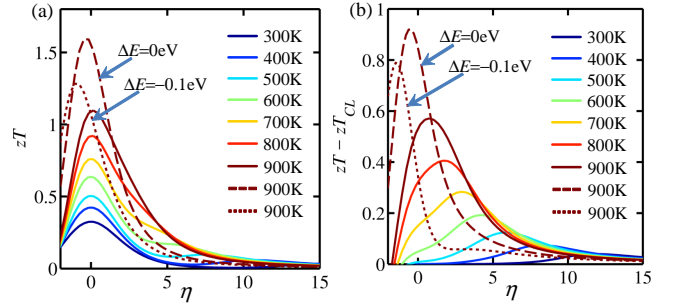

FIG. 3 Calculated (a) $zT$, and (b) $zT - zT_{CL}$ as a function of $\eta$ for SnTe at temperature ranging from 300 – 900 K, respectively.

By definition, $\kappa_{bi}$ is the thermal energy generated by the transition of electrons between different bands [34]. In the three-band case, $\kappa_{bi}$ includes three parts, i.e. the transition of electrons between CB and VB$_L$, between CB and VB$_\Sigma$, and between VB$_L$ and VB$_\Sigma$ [34]. To activate the contribution from VB$_\Sigma$, SnTe should be heavily doped, wherein the transitions of electrons between CB and VB$_L$, and between CB and VB$_\Sigma$ are quite weak [37]. In this regard, $\kappa_{bi}$ is dominated by transition of electrons between VB$_L$ and VB$_\Sigma$. For small $\Delta E$, the average thermal energy caused by the transition of electrons between VB$_L$ and VB$_\Sigma$ is low. Therefore, we can observe $\kappa_{bi}$ decreases with reducing $\Delta E$ at 900 K.

Based on the calculated $S^2\sigma$, $\kappa_e$, $\kappa_{bi}$, and obtained $\kappa_l$, we calculated $zT$. Fig. 3(a) shows the calculated $zT$ as a function of $\eta$ at temperature ranging between 300 and 900 K. As can be seen, different from the observed two peaks for $S^2\sigma$ in Fig. 3(a), $zT$ has only one peak at a given $T$, which is caused by the increased $\kappa_e$ and $\kappa_{bi}$ at large $\eta$. Moreover, with reducing $\Delta E$ to 0 eV at 900 K, $zT$ for SnTe is predicted to be 1.6, which means SnTe as a Pb-free rock-salt chalcogenide is a promising candidate working at the mid-temperature range.

Fig. 3(b) shows the plots of temperature-dependent $zT - zT_{CL}$ as a function of $\eta$ to clarify the contribution of VB$_\Sigma$, in which $zT - zT_{CL}$ peaks increase with increasing $T$. At 900 K with the inherent $\Delta E$, the $zT$ peak reaches 1.2. Correspondingly, the $zT - zT_{CL} \approx 0.6$ can be obtained at $T = 900$ K from Fig. 3(b), indicating that nearly 50% of peak $zT$ is contributed from VB$_\Sigma$. Moreover, for $\Delta E = 0$ eV at 900 K, the significantly increased peak $zT$ of 1.6 is caused by the increased $zT - zT_{CL}$. Therefore, we can conclude that the contribution of VB$_\Sigma$ is essential in enhancing the overall $zT$, particularly at high temperature.

So far, we illustrated the thermoelectric properties as a function of $\eta$ for SnTe, which describes how the band structure affects the thermoelectric properties, and we found that $zT$ for SnTe reaches up to 1.6 for $\Delta E = 0$ eV at 900 K. Following this, we predicted the maximum $zT$ in both PbTe and PbSe. To achieve the maximum $zT$, one condition is to produce the convergence of VB$_L$ and VB$_\Sigma$ at the highest working temperature, and the other is to properly tune $n_H$. As such, it is necessary to determine the corresponding $n_H^{opt}$ for these rock-salt structured chalcogenides. To this end, we calculated the thermoelectric properties as functions of $T$ and $n_H$ for SnTe, PbTe and PbSe with band convergence occurring at 900 K, shown in Figs. S2 – S4. The feasibility of our calculations was verified via comparing our calculated $S$, $\mu_H$, $S^2\sigma$, and $zT$ as a function of $n_H$ with the reported experimental values of Na-doped PbSe [10], Na-doped PbTe [29], and I-doped SnTe [14]. All the comparisons are shown in Figs. S5 – S7, in which experimental values can be well matched with our calculation curves of $S$, $\mu_H$, $S^2\sigma$, or $zT$ (from CB+VB$_L$+VB$_\Sigma$) at a wide temperature range, confirming the feasibility of our calculations.

On this basis, we determined the $n_H^{opt}$ for $S^2\sigma$, and for $zT$, shown in Figs. 4(a) and 4(b), respectively. As can be seen, $n_H^{opt}$ for $zT$ is lower than that for $S^2\sigma$, which is to compromise the increased $\kappa_e$ at large $n_H$ (i.e. large $\eta$). Moreover, we calculated the temperature dependent $S^2\sigma$ and $zT$ for a given $n_H$. Fig. 4(c) shows the temperature dependent $S^2\sigma$ for various $n_H$ values, in which the dash curves are the temperature dependent $S^2\sigma$ for $n_H = n_H^{opt}$ for $S^2\sigma$ at 900 K, and the solid curves are the temperature dependent $S^2\sigma$ for $n_H = n_H^{opt}$ for $zT$ at 900 K. As can be seen, the solid curves are lower than the corresponding dash curves, which means at $n_H^{opt}$ for maximizing $zT$, the corresponding $S^2\sigma$ is not maximized. Fig. 4(d) shows the temperature dependent $zT$ (solid curves) for $n_H$ equaling to the 900 K $n_H^{opt}$ with the reported $\kappa_l$ from Ref. [29] for PbTe, Ref. [10] for PbSe, and Ref. [14] for SnTe. As can be seen, the maximum $zT$ values for PbTe, PbSe, and SnTe are predicted to be 2.2, 1.8, and 1.6. In addition, if we used the amorphous limit $\kappa_l$ of 0.36 Wm$^{-1}$K$^{-1}$ for PbTe [38], of 0.35 Wm$^{-1}$K$^{-1}$ for PbSe [39], of 0.4

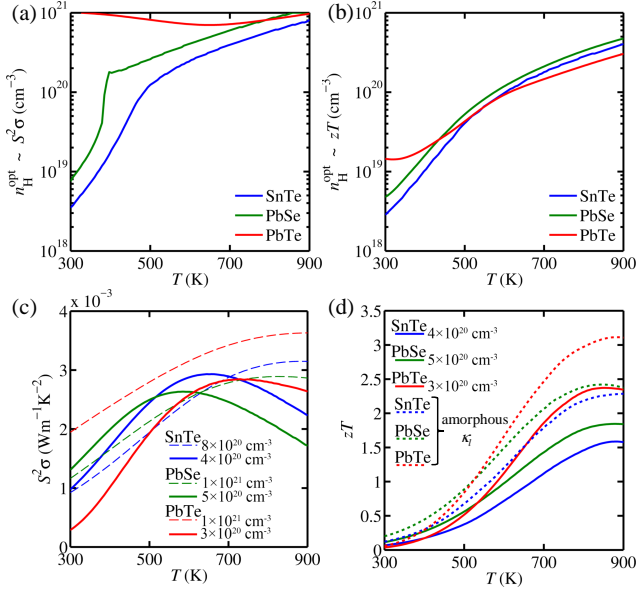

FIG. 4 Determined temperature-dependent $n_H^{opt}$ for (a) $S^2\sigma$ and (b) $zT$. (c) Calculated $S^2\sigma$ and (d) calculated $zT$ as a temperature for SnTe, PbSe, and PbTe.

Wm$^{-1}$K$^{-1}$ for SnTe [40], the maximum $zT$ could be further enhanced to 3.1, 2.4, and 2.2, respectively.

In this study, we performed simulations based on the three bands model for SnTe as an example. Qualitatively, we found that along with the enhancement in $S^2\sigma$ caused by VB$_\Sigma$, $\kappa_e$ and $\kappa_{bi}$ also increases, due to the extra transition of electrons between VB$_L$ and VB$_\Sigma$. This can partially offset the enhancement in $S^2\sigma$ caused by VB$_\Sigma$ for the overall $zT$. Moreover, we determined the $n_H^{opt}$ for $S^2\sigma$ and $zT$ in PbTe, PbSe, and SnTe, which suggests that highly doped $p$-type rock-salt structured chalcogenides is required. Combining the reported $\kappa_l$ from literatures, we predicted the maximum $zT$ values for PbTe, PbSe, and SnTe can be 2.2, 1.8, and 1.6, respectively. If $\kappa_l$ reaches to the amorphous limit, $zT$ could be further enhanced to 3.1, 2.4, and 2.2, respectively. This study suggests that there is plenty room for $zT$ improvement in the currently reported PbTe, PbSe, and SnTe when producing the band convergence at 900 K with properly tuned $n_H$.

This work was financially supported by the Australian Research Council. MH thanks the China Scholarship Council for providing the Ph.D. stipend, and the Graduate School of University of Queensland for providing the travel award.